\documentstyle[sprocl,rotate]{article}

\input{psfig}

\newcommand{\rf}[4]{{\em {#1}} {\bf #2}, #3 (#4)}

\newcommand{\pr}{Phys.\ Rev.\ }

\newcommand{\pl}{Phys.\ Lett.\ }

\newcommand{\np}{Nucl.\ Phys.\ }

\def\be{\begin{equation}}
\def\ee{\end{equation}}
\def\bea{\begin{eqnarray}}
\def\eea{\end{eqnarray}}
\def\Tr{{\rm Tr}\,}
\def\muhat{\hat{\mu}}
\def\qhat{\hat{q}}
\def\bra{\langle}
\def\ket{\rangle}
\def\half{\frac{1}{2}}
\def\sss{\scriptscriptstyle}
\def\Muv{M_{\sss\rm UV}}
\def\Mir{M_{\sss\rm IR}}
\newcommand{\err}[2]{\raisebox{-0.4ex}
{$\stackrel{\scriptstyle +#1}{\scriptstyle -#2}$}}

\begin{document}

\title{Lattice gauge theory studies of the gluon propagator}

\author{D.~B.~Leinweber, J.~I.~Skullerud,\footnote{UKQCD
Collaboration} and \protect\underline{A. G. Williams}}

\address{CSSM {\em and} Department of Physics and Mathematical Physics,\\
The University of Adelaide, SA 5005, Australia\\
http://www.physics.adelaide.edu.au/theory/}

\maketitle\abstracts{ The gluon propagator in Landau gauge is
calculated in quenched QCD on a large ($32^3\times 64$) lattice at
$\beta=6.0$.  In order to assess finite volume and finite lattice
spacing artefacts, we also calculate the propagator on a smaller
volume for two different values of the lattice spacing.  New structure
seen in the infrared region survives conservative cuts to the lattice
data, and serves to exclude a number of models that have appeared in
the literature.}

\section{Introduction}
\label{sec:intro}

The infrared behaviour of the gluon propagator is important for an
understanding of confinement.  Previous conjectures range from a
strong divergence~\cite{mandelstam,bp} to a propagator that vanishes
in the infrared~\cite{gribov,stingl}.

Lattice QCD should in principle be able to resolve this issue by
first-principles, model-independent calculations.  However, lattice
studies have been inconclusive up to now,\cite{bps,mms}
since they have not been able to access sufficiently low momenta.  The
lower limit of the available momenta on the lattice is given by
$q_{\rm min} = 2\pi/L$, where $L$ is the length of the lattice.
Here we will report results using a lattice with a length of 3.3~fm in
the spatial directions and 6.7~fm in the time direction.  This gives us
access to momenta as small as 400~MeV.

\section{Lattice formalism}
\label{sec:def}

The gluon field $A_\mu$ can be extracted from the link variables
$U_\mu(x)$ using
\be
U_\mu(x) =  e^{ig_0aA_\mu(x+\muhat/2)} + {\cal O}(a^3) \, .
\ee
Inverting and Fourier transforming this, we obtain
\bea
A_\mu(\qhat) & \equiv & \sum_x e^{-i\qhat\cdot(x+\muhat/2)}
 A_\mu(x+\muhat/2) \nonumber \\
 & = & \frac{e^{-i\qhat_{\mu}a/2}}{2ig_0a}\left[\left(U_\mu(\qhat)-U^{\dagger}_\mu(-\qhat)\right)
 - \frac{1}{3}\Tr\left(U_\mu(\qhat)-U^{\dagger}_\mu(-\qhat)\right)\right] , 
\eea
where $U_\mu(\qhat)\equiv\sum_x e^{-i\qhat x}U_\mu(x)$ and
$A_\mu(\qhat)\equiv t^a A_{\mu}^a(\qhat)$.  The available momentum
values $\qhat$ are given by
\be
\qhat_\mu  = 2 \pi n_\mu/(a N_\mu), \qquad
n_\mu=0,\ldots,N_\mu-1
\ee
where $N_\mu$ is the number of points in the $\mu$ direction.  The
gluon propagator $D^{ab}_{\mu\nu}(\qhat)$ is defined as
\be
D^{ab}_{\mu\nu}(\qhat) = \bra A^a_\mu(\qhat)
A^b_\nu(-\qhat) \ket\,/\,V \, .
\ee

\noindent In the continuum Landau gauge, the propagator has the
structure
\be
D_{\mu\nu}^{ab}(q) =
\delta^{ab}(\delta_{\mu\nu}-\frac{q_{\mu}q_{\nu}}{q^2})D(q^2)
\, .
\label{eq:landau-prop}
\ee
At tree level, $D(q^2)$ will have the form
\be
D^{(0)}(q^2) = 1/q^2\, .
\label{eq:tree}
\ee
On the lattice, this becomes
\be
D^{(0)}(\qhat) = 
1/\sum_{\mu}\left(\frac{2}{a}\sin\frac{\qhat_{\mu}a}{2}\right)\, .
\label{eq:lat-tree}
\ee
Since QCD is asymptotically free, we expect that up to logarithmic
corrections, $q^2 D(q^2) \to 1$ in the ultraviolet.  Hence we define
the new momentum variable $q$ by
\be
q_\mu \equiv \frac{2}{a}\sin\frac{\qhat_\mu a}{2} ,
\label{eq:lat-momenta}
\ee
and work with this throughout.

The (bare) lattice gluon propagator is related to the renormalised
continuum propagator $D_R(q;\mu)$ via
\be
D^L(qa) = Z_3(\mu,a) D_R(q;\mu) \, .
\label{eq:renorm-def}
\ee
The renormalisation constant $Z_3(\mu,a)$ can be found by imposing a
momentum subtraction renormalisation condition
$
D_R(q)|_{q^2=\mu^2} = \frac{1}{\mu^2} \, .
\label{eq:renorm-mom}
$

The asymptotic behaviour of the renormalised gluon propagator in the
continuum is given to one-loop level by
\be
D_R(q) \equiv D_{\rm bare}(q)/Z_3
= \frac{1}{q^2}\left(\half\ln(q^2/\Lambda^2)\right)^{-d_D}
\label{model:asymptotic}
\ee
with 
\be
d_D=\frac{39-\xi-4N_f}{4(33-2N_f)}=\frac{13}{44} \, ,
\ee
where both the gauge parameter $\xi$ and
the number of fermion flavours $N_f$ are zero in this calculation.

\section{Simulation Parameters, Finite Size Effects and Anisotropies}
\label{sec:params}

We have analysed three lattices, with different values for the volume
and lattice spacing.  The details are given in table~\ref{tab:sim-params}.
In the following, we are particularly interested in the deviation of
the gluon propagator from the tree level form.  We will therefore
factor out the tree level behaviour and plot $q^2 D(q^2)$ rather than
$D(q^2)$ itself.

\begin{table}[tb]
\begin{center}
\leavevmode
\begin{tabular}{lcccrcc}
\hline
Name &$\beta$ &$a_{\rm st}^{-1}$ (GeV) &Volume &$N_{\rm conf}$ &$a \widehat
q_{\rm Max}$ & $|\partial_\mu A_\mu|$ \\ 
\hline
\hline
Small    &6.0  &1.885   &$16^3\times 48$ &125   &$2 \pi / 4$ & $< 10^{-6}$ \\
Large    &6.0  &1.885   &$32^3\times 64$ &75    &$2 \pi / 4$ & $< 10^{-6}$ \\
Fine     &6.2  &2.63    &$24^3\times 48$ &223   &$2 \pi / 4$ & $< 10^{-6}$ \\
\hline
\end{tabular}
\end{center}
\caption{Simulation parameters}
\label{tab:sim-params}
\end{table}

\begin{figure}[bt]
\begin{center}
\leavevmode
\mbox{\rotate[l]{\psfig{figure=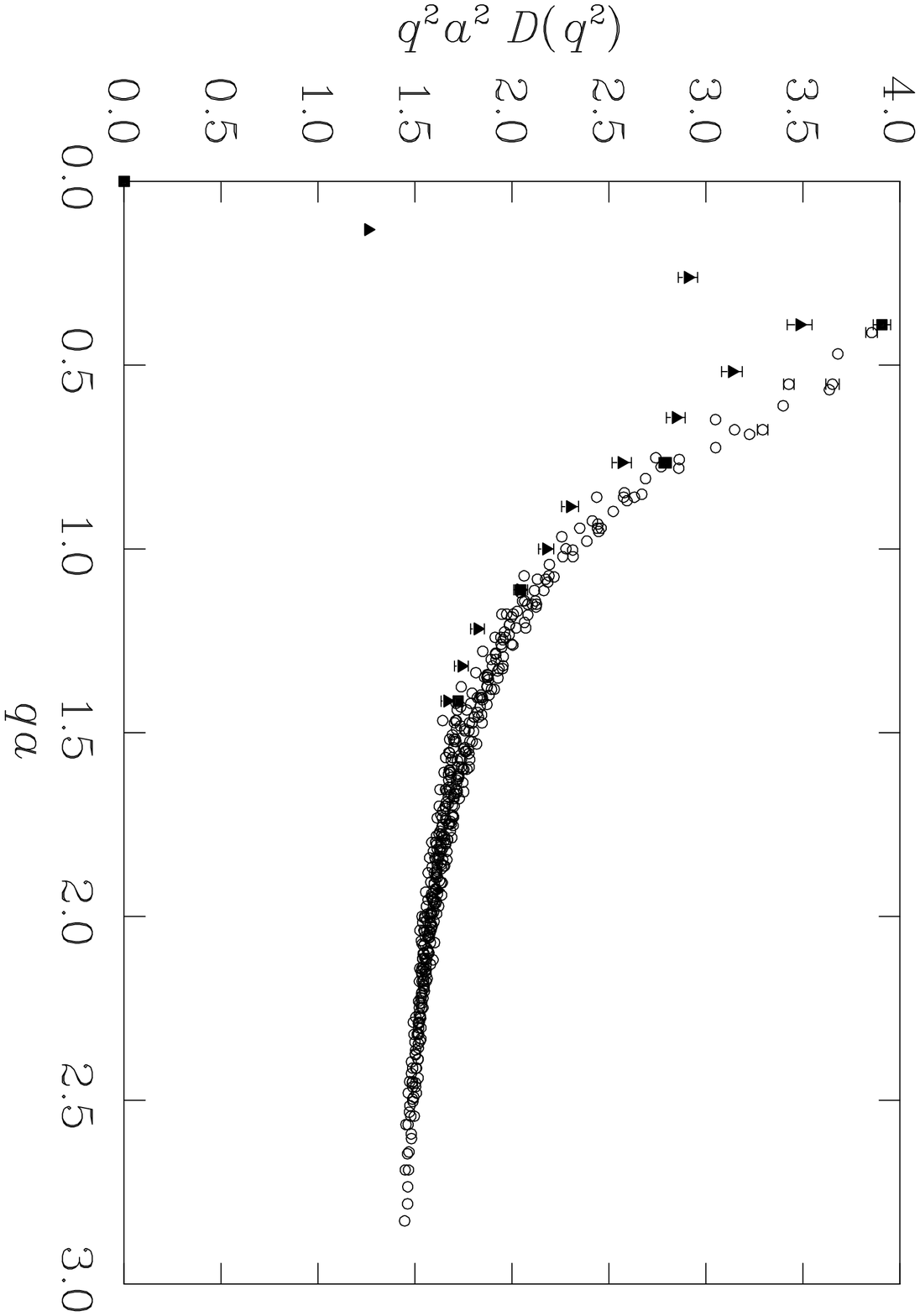,height=2.2in}}\hspace{0.5cm}
\rotate[l]{\psfig{figure=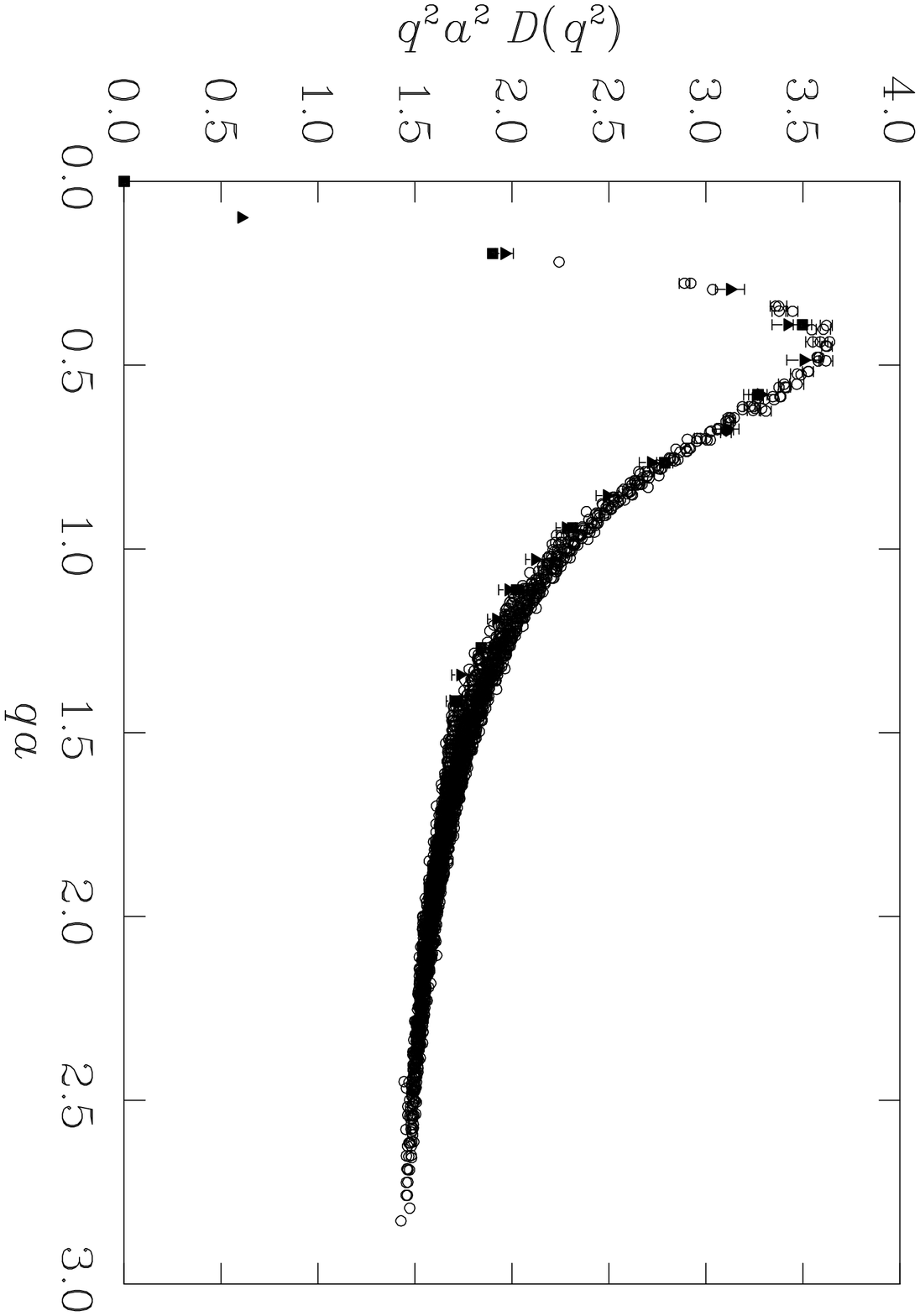,height=2.2in}}}
\end{center}
\caption{Componentwise data, for the small lattice (left) and the
large lattice (right).  The filled triangles denote momenta directed
along the time axis, while the filled squares denote momenta directed
along one of the spatial axes.}
\label{fig:cpt-data}
\end{figure}

Fig.~\ref{fig:cpt-data} shows the gluon propagator as a function of
$qa$ for the small and large lattices, with momenta in different
directions plotted separately.  For low momentum values on the small
lattice, there are large discrepancies due to finite size effects
between points representing momenta along the time axis and those
representing momenta along the spatial axes.  These discrepancies are
absent from the data from the large lattice, indicating that finite
size effects here are under control.

However, at higher momenta, there are anisotropies which remain for
the large lattice data, and which are of approximately the same
magnitude for the two lattices.  In order to eliminate these
anisotropies, which arise from finite lattice spacing errors, we
select momenta lying within a cylinder of radius $\Delta\qhat a =
2\times 2\pi/32$ along the 4-dimensional diagonals.\cite{letter}

\section{Scaling behaviour}

Since the renormalised propagator $D_R(q;\mu)$ is independent of the
lattice spacing, we can use (\ref{eq:renorm-def}) to derive a simple,
$q$-independent expression for the ratio of the unrenormalised
lattice gluon propagators at the same value of $q$:
\be
\frac{D_c(qa_c)}{D_f(qa_f)} = 
\frac{Z_3(\mu,a_c)D_R(q;\mu)/a_c^2}{Z_3(\mu,a_f)D_R(q;\mu)/a_f^2}
= \frac{Z_c}{Z_f}\frac{a_f^2}{a_c^2}
\label{eq:gluon_match_ratio}
\ee
where the subscript $f$ denotes the finer lattice ($\beta=6.2$ in this
study) and the subscript $c$ denotes the coarser lattice
($\beta=6.0$).  We can use this relation to study directly the scaling
properties of the lattice gluon propagator by matching the data for
the two values of $\beta$.  This matching can be performed by
adjusting the values for the ratios $R_Z = Z_f/Z_c$ and $R_a =
a_f/a_c$ until the two sets of data lie on the same curve.

We have implemented this by making a linear interpolation of the
logarithm of the data plotted against the logarithm of the momentum
for both data sets.  In this way the scaling of the momentum is
accounted for by shifting the fine lattice data to the right by an
amount $\Delta_a$ as follows
\be
\ln D_c( \ln(qa_c) ) = \ln D_f( \ln(qa_c) - \Delta_a ) + \Delta_Z
\label{eq:match-data}
\ee
Here $\Delta_Z$ is the amount by which the fine lattice data must be
shifted up to provide the optimal overlap between the two data sets.
The matching of the two data sets has been performed for values of
$\Delta_a$ separated by a step size of 0.001.  $\Delta_Z$ is
determined for each value of $\Delta_a$ considered, and the optimal
combination of shifts is identified by searching for the global
minimum of the $\chi^2/$dof.  The ratios $R_a$ and $R_Z$ are related
to $\Delta_a$ and $\Delta_Z$ by
\be
R_a = e^{-\Delta_a}\, , \hspace{2.0em} R_Z = R_a^2 e^{-\Delta_Z} \, .
\ee

\begin{figure}[t]
\begin{center}
\leavevmode
\mbox{\rotate[l]{\psfig{figure=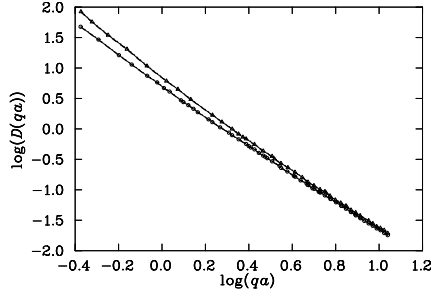,height=2.2in}}}
\end{center}
\caption{The dimensionless, unrenormalised gluon propagator as a
function of $\ln(qa)$ for the two values of $\beta$.  The triangles
denote the data for the small lattice at $\beta=6.0$, while the
circles denote the data for $\beta=6.2$.}
\label{fig:compare_data_lattmom}
\end{figure}

\begin{figure}[tbp]
\begin{center}
\leavevmode
\mbox{\rotate[l]{\psfig{figure=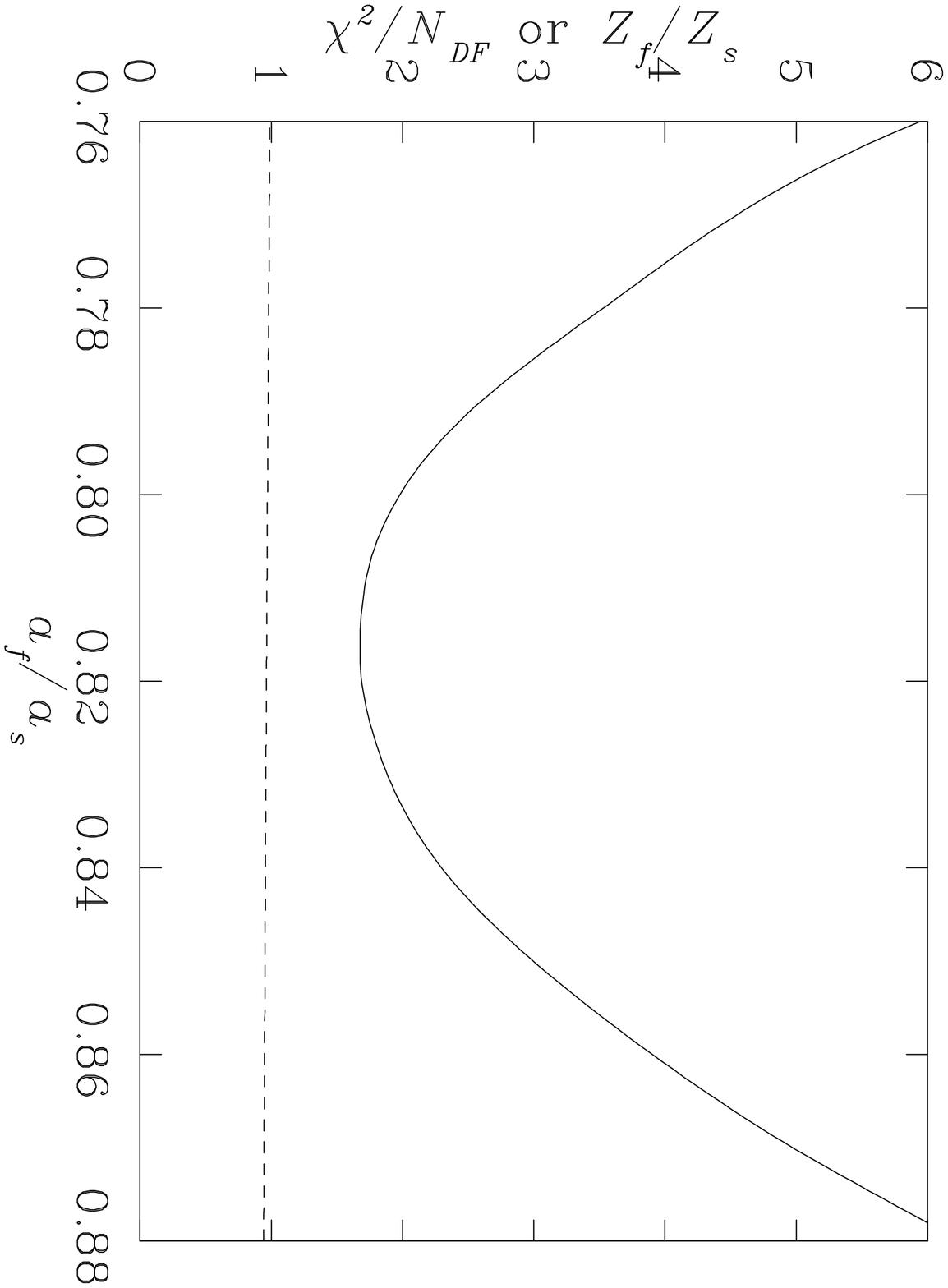,height=2.2in}}\hspace{0.5cm}
\rotate[l]{\psfig{figure=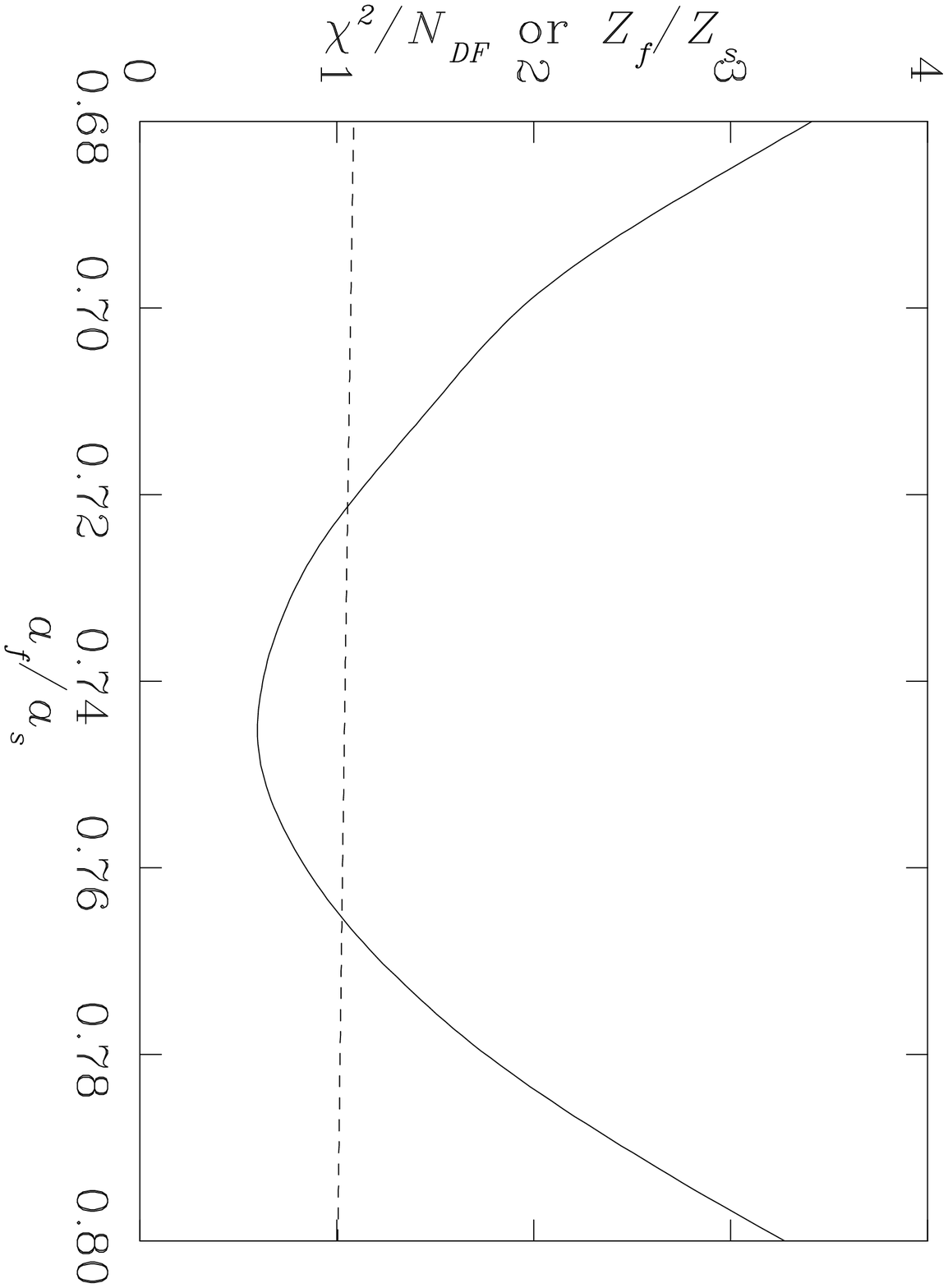,height=2.2in}}}
\end{center}
\caption{$\chi^2$ per degree of freedom as a function of the ratio of
lattice spacings for matching the small and fine lattice data
according to (\protect\ref{eq:match-data}), using $\qhat$ (left) and
$q$ (right) as the momentum variable.  The dashed line indicates the
ratio $R_Z$ of the renormalisation constants.}
\label{fig:match_spacing}
\end{figure}

Fig.~\ref{fig:compare_data_lattmom} shows the data for both lattice
spacings as a function of $qa$ before shifting.  In the leftmost plot of
fig.~\ref{fig:match_spacing} we present the result of the
matching using $\qhat$ as the momentum variable.  The minimum value
for $\chi^2/$dof of about 1.7 is obtained for $R_a\sim 0.815$.  This
value for $R_a$ is considerably higher than the value of $0.716\pm
0.040$ obtained from an analysis of the static quark potential in
\cite{bs}.  From this discrepancy, as well as the relatively high
value for $\chi^2/$dof, we may conclude that the gluon propagator,
taken as a function of $\qhat$, does not exhibit scaling behaviour for
the values of $\beta$ considered here.

The rightmost plot of fig.~\ref{fig:match_spacing} shows the result of
the matching using $q$ of (\ref{eq:lat-momenta}) as the momentum
variable.  We can see immediately that this gives much more
satisfactory values both for $\chi^2/$dof and for $R_a$.  The minimum
value for $\chi^2/$dof of 0.6 is obtained for $R_a=0.745$.  Taking a
confidence interval where $\chi^2/{\rm dof} < \chi^2_{min} + 1$ gives
us an estimate of $R_a=0.745\err{32}{37}$, which is fully compatible
with the value~\cite{bs} of $0.716\pm 0.040$.  The corresponding
estimate for the ratio of the renormalisation constants is $R_Z =
1.038\err{26}{21}$.

\section{Model Fits}
\label{sec:models}

We have demonstrated scaling in our lattice data over the entire range
of $q^2$ considered, and will now proceed with model fits.
The following functional forms have been considered:
\bea
D(q^2) = & 
\frac{Z q^2}{q^4+M^4}\left(\half\ln\frac{q^2+M^2}{M^2}\right)^{-d_D}
 & \hspace{-0.3cm}\mbox{(Gribov\,\cite{gribov})}
\label{model-first}
\label{model:lita}
\label{model:gribov} \\
D(q^2) = & Z\frac{q^2}{q^4+2A^2q^2+M^4}
\left(\half\ln\frac{q^2+M^2}{M^2}\right)^{-d_D}
 & \hspace{-0.1cm}\mbox{\hfill(Stingl\,\cite{stingl})}
\label{model:stingl} \\
D(q^2) = & \frac{Z}{(q^2)^{1+\alpha}+M^2}
 & \hspace{-1.7cm}\mbox{\hfill(Marenzoni {\em et al}\,\,\cite{mms})}
\label{model:marenzoni} \\
D(q^2) = & Z
\left[(q^2+M^2(q^2))\ln\frac{q^2+4M^2(q^2)}{\Lambda^2}\right]^{-1}
 & \hspace{-0.6cm}\mbox{(Cornwall\,\cite{cornwall})}
\label{model:cornwall} \\
 \mbox{where} & 
M(q^2) = M\left\{\ln\frac{q^2+4M^2}{\Lambda^2}/
\ln\frac{4M^2}{\Lambda^2}\right\}^{-6/11} \nonumber \\
D(q^2) = & Z\left(\frac{A}{(q^2+\Mir^2)^{1+\alpha}} + 
\frac{1}{q^2+\Muv^2}\left(\half\ln\frac{q^2+\Muv^2}{\Muv^2}\right)^{-d_D}\right)
\label{modelA} \\
D(q^2) = & Z\left(\frac{A}{(q^2)^{1+\alpha}+(\Mir^2)^{1+\alpha}} +
\frac{1}{q^2+\Muv^2}\left(\half\ln\frac{q^2+\Muv^2}{\Muv^2}\right)^{-d_D}\right) 
\label{modelB} \\
D(q^2) = & Z\left(A e^{-(q^2/\Mir^2)^{\alpha}} + 
\frac{1}{q^2+\Muv^2}\left(\half\ln\frac{q^2+\Muv^2}{\Muv^2}\right)^{-d_D}\right)
\label{modelC}
\label{model-last}
\eea

We have also considered~\cite{next} special cases of the three forms
(\ref{modelA})--(\ref{modelC}), with $\Mir=\Muv$ or with specific
values for the exponent $\alpha$.  Equations (\ref{model:gribov}) and
(\ref{model:stingl}) are modified in order to
exhibit the asymptotic behaviour of (\ref{model:asymptotic}).
Models~(\ref{modelA}) and (\ref{modelB})
are constructed as generalisations of (\ref{model:marenzoni}) with the
correct dimension and asymptotic behaviour.

All models were fitted to the large lattice data using the cylindrical
cut.  The lowest momentum value was excluded, as the volume dependence
of this point could not be assessed.  In order to balance the
sensitivity of the fit between the high- and low-momentum region,
nearby data points within $\Delta(qa) < 0.05$ were averaged.

\begin{figure}[tb]
\begin{center}
\leavevmode
\mbox{\rotate[l]{\psfig{figure=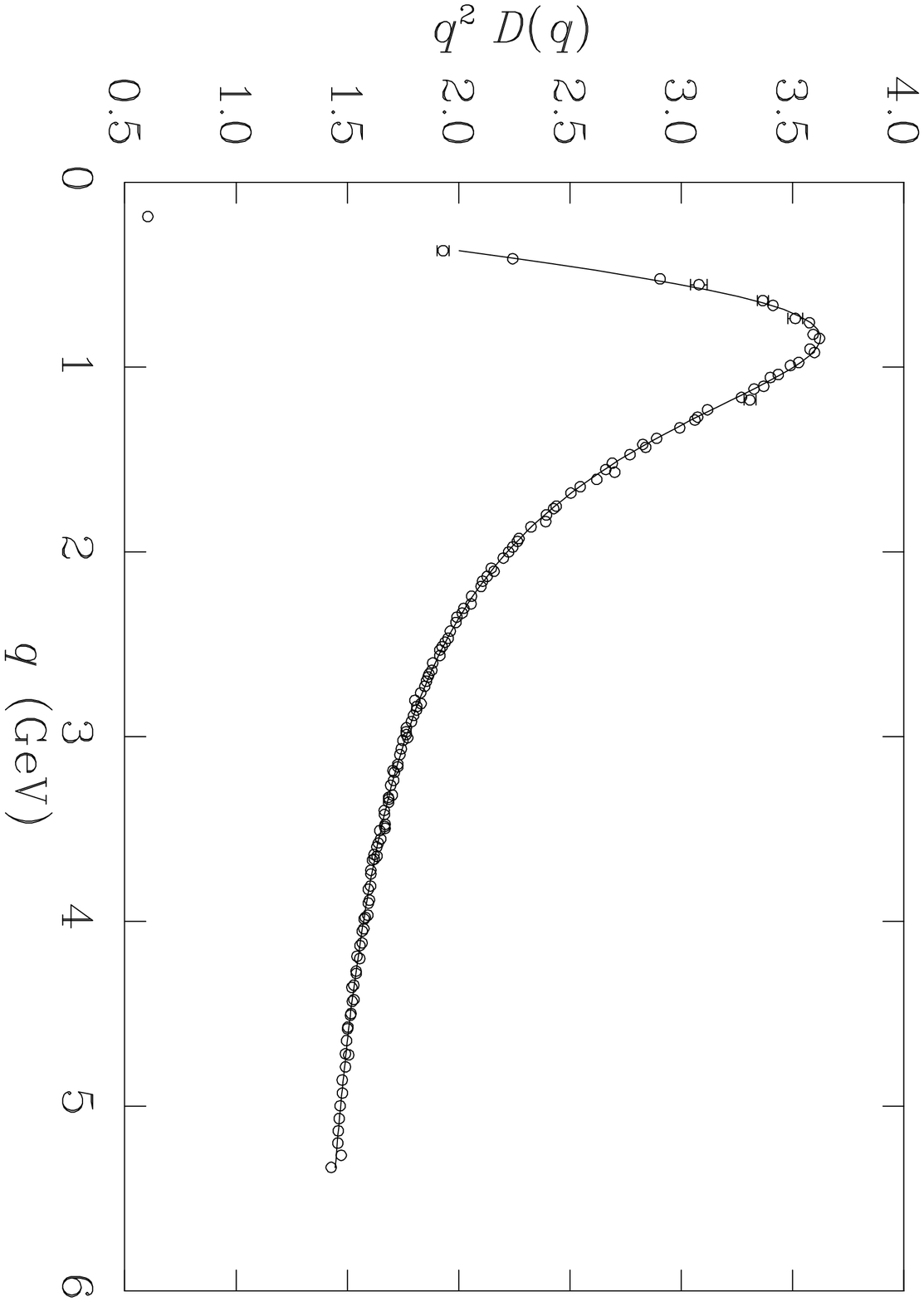,height=2.2in}}}
\end{center}
\caption{The gluon propagator multiplied by $q^2$, with nearby points
averaged.  The line illustrates our best fit to the form defined in
(\protect\ref{modelB}).  The fit is performed over all points shown,
excluding the one at the lowest momentum value, which may be sensitive
to the finite volume of the lattice.  The scale is taken from the
string tension~\protect\cite{bs}.  }
\label{fig:fit-modelB}
\end{figure}

$\chi^2$ per degree of freedom and parameter values for fits to all
these models are shown in table~\ref{tab:fit-params}.  It is clear
that model (\ref{modelB}) accounts for the data better than any of the
other models.  The best fit to this model is illustrated in
fig.~\ref{fig:fit-modelB}.

\begin{table}
\begin{tabular}{c|c|ccccc}
Model & $\chi^2/$dof & $Z$ & $A$ & $\Mir$ & $\Muv$ or $\Lambda$ & $\alpha$
\\ \hline
\ref{model:gribov} & 1972 & 2.19\err{31}{15} & & 0.23\err{14}{1} \\
\ref{model:stingl} & 1998 & 2.2 & 0 & 0.23 \\
\ref{model:marenzoni} & 163 & 2.41\err{0}{12} & & 0.14\err{4}{14}
 & & 0.29\err{6}{2} \\
\ref{model:cornwall} & 50.3 & 6.5\err{7}{9} & & 0.24\err{3}{16} &
0.27\err{7}{7}  \\
\ref{modelA} & 2.96 & 1.54\err{10}{20} & 1.24\err{21}{21}
 & 0.46\err{2}{14} & 0.96\err{47}{17} & 1.31\err{16}{43} \\
\ref{modelA}; $\Mir\!=\!\Muv$ & 3.73 & 1.71\err{9}{0} & 0.84\err{0}{29} & 0.48\err{2}{17}
 & & 1.52\err{12}{37} \\
\ref{modelB} & 1.57 & 1.78\err{45}{20} & 0.49\err{17}{6}
 & 0.43\err{5}{1} & 0.20\err{37}{19} & 0.95\err{7}{5} \\
\ref{modelB}; $\Mir\!=\!\Muv$ & 4.00 & 1.62\err{3}{4} & 0.58\err{5}{1} & 0.40\err{9}{2}
 & & 0.92\err{17}{1} 
\\
\ref{modelC} & 47 & 2.09\err{30}{12} & 29\err{166}{2}
 & 0.22\err{0}{16} & 0.14\err{11}{10} & 0.49\err{0}{16}
\end{tabular}
\caption{Parameter values for fits to models
(\protect\ref{model-first})--(\protect\ref{model-last}).  The
values quoted are for fits to the entire set of data.  The errors
denote the uncertainties in the last digit(s) of the parameter values
which result from varying the fitting range.  The fitting ranges
considered when evaluating the uncertainties are those with a minimum
of 40 points included and with the minimum value for $qa$ no larger
than 1.03 (point number 40).}
\label{tab:fit-params}
\end{table}

\section{Discussion and Outlook}
\label{sec:discuss}

We have evaluated the gluon propagator on an asymmetric lattice with a
large physical volume.  By studying the anisotropies in the data, and
comparing the data with those from a smaller lattice, we have been
able to conclude that finite size effects are under control on the
large lattice.

A clear turnover in the behaviour of $q^2 D(q^2)$ has been observed at
$q \sim 1$GeV, indicating that the gluon propagator diverges less
rapidly than $1/q^2$ in the infrared, and may be infrared finite or
vanishing. 

The data are consistent with a functional form $D(q^2) = D_{\rm IR} +
D_{\rm UV}$, where
\be
D_{\rm IR} = \half \frac{1}{q^4+M^4} ,
\label{eq:ir-prop}
\ee
$M \sim 1$ GeV, and $D_{\rm UV}$ is the asymptotic form given by
(\ref{model:asymptotic}).  A more detailed analysis~\cite{next} of the
asymptotic behaviour reveals that the one-loop formula
(\ref{model:asymptotic}) remains insufficient at $q^2=50{\rm GeV}^2$.

Issues for future study include
the effect of Gribov copies and of dynamical fermions.
We also hope to use improved actions to
perform realistic simulations at larger lattice spacings.  This would
enable us to evaluate the gluon propagator on larger physical volumes,
giving access to lower momentum values.

\section*{Acknowledgments} 

The numerical work was mainly performed on a Cray T3D at EPCC,
University of Edinburgh, using UKQCD Collaboration time under PPARC
Grant GR/K41663.  Financial support from the Australian Research
Council is gratefully acknowledged.  We thank Claudio
Parrinello for stimulating discussions.


\begin{thebibliography}{99}

\bibitem{mandelstam} S.~Mandelstam, \rf{\pr}{D 20}{3223}{1979}

\bibitem{bp} N.~Brown and M.R.~Pennington, \rf{\pr}{D 39}{2723}{1989}

\bibitem{gribov} V.N.~Gribov, \rf{\np}{B 139}{19}{1978}; D.~Zwanziger,
\rf{\np}{B 378}{525}{1992}

\bibitem{stingl} M.~Stingl, \rf{\pr}{D 34}{3863}{1986}; 
\rf{\pr}{D 36}{651}{1987} 

\bibitem{bps} C.~Bernard, C.~Parrinello, A.~Soni,
\rf{\pr}{D49}{1585}{1994} 

\bibitem{mms} P.~Marenzoni, G.~Martinelli, N.~Stella, 
\rf{\np}{B 455}{339}{1995}; P.~Marenzoni {\em et al}, 
\rf{\pl}{B 318}{511}{1993}

\bibitem{cornwall} J.~Cornwall, \rf{\pr}{D 26}{1453}{1982}

\bibitem{bs} G.S.~Bali and K.~Schilling, \rf{\pr}{D 47}{661}{1993}

\bibitem{letter} D.B.~Leinweber, C.~Parrinello, J.I.~Skullerud,
A.G.~Williams, \rf{\pr}{D 58}{031501}{1998}

\bibitem{next} D.B.~Leinweber, C.~Parrinello,
J.I.~Skullerud, A.G.~Williams, in preparation.

\end{thebibliography}
\end{document}